\documentstyle[aps,prb,preprint]{revtex}
\begin{document}
\draft
\title{Non-Perturbative Approach to Nearly Antiferromagnetic Fermi Liquids}
\author{P. Monthoux}
\address{Cavendish Laboratory, University of Cambridge
\\Madingley Road, Cambridge, CB3 0HE, UK}
\date{\today}
\maketitle

\begin{abstract}

We present a non-perturbative approach to the problem of quasiparticles 
coupled to spin-fluctuations. If the fully dressed spin-fluctuation propagator
is used in the Feynman graph expansion of the single particle Green's function,
the problem of summing all spin-fluctuation exchange Feynman graphs (i.e
all the graphs without fermion loops) can be cast as a functional integral 
over Gaussian distributed random vector fields. The quasiparticle propagator 
is then obtained as the average over all field configurations of the 
non-interacting electron Green's function in an external field. A Monte Carlo 
sampling of this functional integral does not suffer from the "fermion sign 
problem" and offers an attractive alternative to perturbative calculations. 
We compare the results of our computer simulations to perturbation theory 
and self-consistent Eliashberg calculations.

\end{abstract}

\vfill\eject

\narrowtext

\section{Introduction}

In previous papers (Monthoux and Pines, 1993, 1994) we described the results 
of extensive one-loop self-consistent calculations on the normal state 
and superconducting transition temperature for the optimally doped copper
oxide superconductor $\rm YBa_2Cu_3O_7$. In these calculations, the effective 
low-energy quasiparticle-quasiparticle interaction is expressed as the 
exchange of dressed magnetic excitations whose spectrum is determined by 
fits to NMR experiments. These Eliashberg calculations demonstrated that 
while strong coupling effects reduce $\rm T_c$ substantially, $\rm d_{x^2-y^2}$
superconductivity at high temperatures is still possible with a 
quasiparticle-spin-fluctuation coupling constant $\rm g^2$ in the range 
required to explain the observed normal-state resisitivity and optical
conductivity.

Questions regarding the validity of the single spin-fluctuation exchange 
picture have been raised (Schrieffer, 1995). The role of vertex corrections in 
the paramagnetic phase has been examined in detail by Chubukov et al. (1997) 
and the author (1997). For a large Fermi surface consistent with ARPES 
experiments on $\rm YBa_2Cu_3O_7$, vertex corrections enhance the effective 
spin-fluctuation interaction as the antiferromagnetic correlation length 
increases. The effect is strongly anisotropic and most important for 
quasiparticles at the hot spots (Hlubina and Rice, 1997) for which the Midgal 
argument does not carry through. A quasiparticle near the hot spot can emit 
or absorb any number of spin-fluctuations at the antiferromagnetic wavevector 
$\rm \vec Q = (\pi,\pi)$ where the interaction is strong while staying 
close to the Fermi surface. One would thus expect that as the coupling 
gets stronger or as the correlation length increases, two, three, etc.. 
spin-fluctuation exchange graphs become important and perturbation theory 
breaks down. Non-perturbative methods are thus needed to study the behaviour 
of hot quasiparticles for long correlation lengths. In the case where 
the charactersitic spin-fluctuation frequency $\rm \omega_{SF}$ is much 
smaller than $\rm 2\pi T$, only the zero Matsubara frequency of the 
spin-fluctuation propagator is relevant and a solution to the problem in 
this quasi-static limit has been proposed by Schmalian et al (1998).

In this paper, the calculation of the single particle Green's function for
an arbitrary number of spin-fluctuation exchanges is formulated as the
average of the non-interacting electron Green's function in an external 
magnetic field whose probability distribution is Gaussian. Upon expanding 
the Green's function in powers of the magnetic field and doing the averages 
with Wick's theorem, one recovers the standard time-ordered perturbation series
for the quasiparticle propagator. However, sampling the functional
integral by the Monte Carlo method combined with a numerical calculation of
the Green's function for each external field configuration
provides an attractive and non-perturbative method of solution when 
straightforward perturbation theory is expected to break down.

The low-energy effective action for the planar quasiparticles is given by

\begin{eqnarray}
\rm S_{eff} & = & \rm \sum_{\vec p,\alpha}\int_0^\beta d\tau 
\psi^\dagger_{\vec p,\alpha}(\tau)\Big(\partial_\tau + 
\epsilon_{\vec p} - \mu\Big) \psi_{\vec p,\alpha}(\tau) \nonumber\\
& &\rm - {g^2\over 6}\sum_{\vec q}\int_0^\beta 
d\tau \int_0^\beta d\tau' \chi(\vec q,\tau-\tau')
\vec s(\vec q,\tau)\cdot \vec s(-\vec q,\tau')
\label{Seff}
\end{eqnarray}

\noindent The spin density $\rm \vec s(\vec q,\tau)$ is given by

\begin{equation}
\rm \vec s(\vec q,\tau) \equiv \sum_{\vec p,\alpha,\gamma} 
\psi^\dagger_{\vec p + \vec q,\alpha}(\tau)\vec \sigma_{\alpha,\gamma}
\psi_{\vec p,\gamma}(\tau)
\label{Spin}
\end{equation}

\noindent where $\rm \vec \sigma$ denotes the three Pauli matrices. The 
quasiparticle dispersion relation is

\begin{equation}
\rm \epsilon_{\vec p} = -2t(cos(p_x) + cos(p_y)) - 4t'cos(p_x)cos(p_y)
\label{eps}
\end{equation}

\noindent with hopping matrix elements t=0.25 eV and t'=-0.45t, as suggested
by Si et al. (1993) from fits to band theory and ARPES experiments. We use
units in which the lattice constant is unity. $\mu$ denotes the chemical 
potential, $\beta$ the inverse temperature, $\rm g^2$ the coupling constant
and $\rm \psi^\dagger_{\vec p,\sigma}$ and $\rm \psi_{\vec p,\sigma}$ are
Grassmann variables. A coupling constant $\rm g^2 = 0.41\; eV^2$ gives 
an Eliashberg $\rm T_c$ to a $\rm d_{x^2-y^2}$ superconducting state of 
$\rm 90^\circ K$ (Monthoux and Pines (1994)). Since the interaction 
$\rm \chi(\vec q,\tau)$ is determined from fits to normal state experiments, 
it is fully dressed. Therefore, in a Feynman graph expansion, the diagrams 
corresponding to a renormalization of the spin-fluctuation propagator 
$\rm \chi(\vec q,\tau)$ should not be included. On the real frequency axis, 
$\rm \chi(\vec q,\omega)$ is assumed to take the phenomenological form 
introduced by Millis et al.(1990)

\begin{equation}
\rm \chi(\vec q,\omega) = {\chi_{\vec Q}\over 1 + \xi^2(\vec q - \vec Q)^2
- i{\omega\over\omega_{SF}}}
\label{MMP}
\end{equation}

\noindent Here $\rm \chi_{\vec Q}$ is the static spin susceptibility at 
wavevector $\rm \vec Q = (\pi,\pi)$ and $\xi$ is a temperature dependent
antiferromagnetic correlation length. The frequency $\rm \omega_{SF}$ defines 
the charactersitic energy of the low-frequency magnetic response. The
quantities $\rm \chi_{\vec Q}$ and $\rm \omega_{SF}$ may be expressed in terms
of the experimentally measured long-wavelength spin susceptibility $\rm \chi_0$
and a short-distance characteristic spin-fluctuation energy $\rm \Gamma_{AF}$
by introducing a scale factor $\beta$ (not to be confused with the inverse 
temperature)

\begin{equation}
\rm \chi_{\vec Q} = \chi_0\sqrt{\beta}\xi^2
\label{chiQ}
\end{equation}

\begin{equation}
\rm \omega_{SF} = {\Gamma_{AF}\over \sqrt{\beta}\pi\xi^2}
\label{omsf_def}
\end{equation}

\noindent We adopt the same values of these parameters as in Monthoux and 
Pines (1994), namely $\rm \Gamma_{AF} = 1.3\; eV$, 
$\rm \chi_0 = 2.6\; states/eV$, $\rm \beta = 32$. The temperature dependence
of the charactersitic spin-fluctuation energy $\rm \omega_{SF}$ is taken as

\begin{equation}
\rm \omega_{SF}(T) = (9.5 + 4.75[T(^\circ K)/100]) \; meV
\label{omsf_T}
\end{equation}

\noindent from which the temperature-dependent antiferromagnetic correlation 
length can be obtained using Eq.~(\ref{omsf_def}). The numerical values of 
these  parameters for $\rm YBa_2Cu_3O_7$ are 
$\rm \chi_{\vec Q}(T_c) \approx 75\; states/eV$, 
$\rm \xi(T_c)  \approx 2.3$ and $\rm \omega_{SF}(T_c) \approx 14\; meV$.

The spin-fluctuation propagator on the imaginary axis, 
$\rm \chi(\vec q,i\nu_n)$ is related to the imaginary part of the 
response function $\rm Im\chi(\vec q,\omega)$, Eq.~(\ref{MMP}), via the 
spectral representation

\begin{equation}
\rm \chi(\vec q,i\nu_n) = -\int_{-\infty}^{+\infty}{d\omega\over \pi}
{Im\chi(\vec q,\omega)\over i\nu_n - \omega}
\label{chi_mats}
\end{equation}

\noindent To get $\rm \chi(\vec q,i\nu_n)$ to decay as $\rm 1/\nu_n^2$ as
$\rm \nu_n \rightarrow \infty$, as it should, we introduce a cutoff 
$\rm \omega_0$ and take $\rm Im\chi(\vec q,\omega) = 0$ for $\rm \omega 
\geq \omega_0$. In the following, we adopt the value $\rm \omega_0 
= 0.4\; eV$.

\section{Functional Integral Formulation}

In this section, we formulate the problem of calculating the single particle
Green's function for an arbitrary number of spin-fluctuation exchanges 
as a functional integral that is suitable for computer simulation.\hfil\break

Let us first introduce some notation. In the following, we denote by x the 
position and imaginary time coordinate, $\rm x\equiv (\vec x,\tau)$ and 
similarly, let q stand for the wavevector and Matsubara frequency 3-vector
$\rm q\equiv (\vec q,i\nu_n)$. Consider a real vector field $\rm \vec M(x)$,
such that 

\begin{equation}
\rm \vec M(-q) = \vec M^*(q)
\label{realM}
\end{equation}

\noindent and introduce the average $\rm <...>$ over the Gaussian probability 
distribution of the fields $\rm \vec M(q)$:

\begin{equation}
\rm <....> \equiv {1\over Z} \int \int D\vec M \; .... \;exp\Bigg(-\sum_{q} 
{\vec M(q)\cdot \vec M(-q)\over 2\alpha(q)}\Bigg)
\label{average}
\end{equation}

\noindent where $\rm \alpha(q)$ is real and the normalization factor Z is 
defined as

\begin{equation}
\rm Z = \int \int D\vec M \; exp\Bigg(-\sum_{q} 
{\vec M(q)\cdot \vec M(-q)\over 2\alpha(q)}\Bigg)
\label{PartFunc}
\end{equation}

\noindent The integrals $\rm D\vec M$ are carried out over all of the 
independent components of the vector field $\rm \vec M(q)$. The two-point 
correlation function of the field $\rm \vec M(q)$ is straightforward to
calculate and is given by

\begin{equation}
\rm < M_i(q) M_j(k) > = \delta_{i,j}\delta_{k,-q} \times \Bigg\{ { {2\alpha(q) 
\; if \; M(q) \; is \; complex}\atop {\alpha(q) \; if \; M(q) \; is \; real}}
\label{2ptCorr}
\end{equation}

\noindent Since the integrals over $\rm \vec M(q)$ are Gaussian, the average 
of an odd number of fields vanishes and the higher order correlations can be 
obtained from the two-point correlation function by Wick's theorem. For 
instance the four-point correlation function is given by:

\begin{eqnarray}
\rm < M_i(q) M_j(k) M_n(p) M_m(l) > & = &
\rm < M_i(q) M_j(k) >< M_n(p) M_m(l) > \nonumber\\ 
&  &\rm + < M_i(q) M_n(p) >< M_j(k) M_m(l)> \nonumber\\ 
&  &\rm + < M_i(q) M_m(l) >< M_j(k) M_n(p) >
\label{Wick}
\end{eqnarray}

\noindent Now we choose $\rm \alpha(q)$ such that 

\begin{equation}
\rm < M_i(q)M_i(-q) > = {g^2\over 3}{T\over N}\chi(q)
\label{alpha}
\end{equation}

\noindent where N is the number of momenta in the Brillouin Zone, T the
temperature and let

\begin{equation}
\rm {\tilde G} \equiv < {\tilde G_0}
[1 - {\tilde M} {\tilde G_0}]^{-1} >
\label{Green}
\end{equation}

\noindent where the matrices $\rm {\tilde G_0}$ and $\rm {\tilde M}$ in 
3-momentum and spin space are defined as

\begin{equation}
\rm [{\tilde G_0}]_{(q,\alpha)(q',\beta)} = G_0(q) 
\delta_{(q,\alpha),(q',\beta)}
\label{Gdef}
\end{equation}

\begin{equation}
\rm [{\tilde M}]_{(q,\alpha)(q',\beta)} = \vec \sigma_{\alpha,\beta}\cdot 
\vec M(q-q')
\label{Mdef}
\end{equation}

\noindent where $\rm \vec \sigma$ are the three Pauli matrices and
$\alpha$, $\beta$ spin indices. Upon expanding and doing the averages 
over the random fields $\rm \vec M$, one generates the following 
perturbation expansion:

\begin{eqnarray}
\rm {\tilde G} & \equiv & \rm< {\tilde G_0}
[1 - {\tilde M} {\tilde G_0}]^{-1} >\nonumber\\ 
&  &\rm = {\tilde G_0} + <{\tilde G_0} {\tilde M} 
{\tilde G_0} {\tilde M} {\tilde G_0}> \nonumber\\ 
&  &\rm + <{\tilde G_0} {\tilde M} {\tilde G_0} {\tilde M} 
{\tilde G_0} {\tilde M} {\tilde G_0} {\tilde M} {\tilde G_0}> + \dots
\label{pert}
\end{eqnarray}

\noindent The Feynman graph expansion is shown in fig.(1). One can easily 
convince oneself that the above formulation of the problem amounts to summing 
all spin-fluctuation exchange graphs. Note that no fermion loops are included.
The second order dressing of the spin-fluctuation propagator and higher
order RPA-like diagrams can be absorbed in the definition of the
magnetic response function $\rm \chi(q)$. Higher order fermion loops,
such as the fourth order term with four spin-fluctuation lines 
(non-linear magnetic response), etc.. are not included in the above 
formulation of the problem. These are the terms that give rise to 
the "fermion sign problem" in determinant Monte Carlo simulations. 
Some of these diagrams are shown in fig.(2). The quartic term should in 
principle be introduced in the probability distribution of the vector 
field $\rm \vec M(q)$ when the coefficient $\rm \chi^{-1}(q)$ of the 
quadratic term becomes very small, i.e. near the phase transition. 
In this paper, however, we shall only consider a Gaussian distribution 
of the fields $\rm \vec M(q)$.\hfil

The above formulation immediately suggests a non-perturbative approach to the
problem: instead of expanding the matrix inverse and doing the integrals 
analytically, one can sample the functional integral by the Monte Carlo method
and numerically invert the Green's function matrix in Eq.~(\ref{Green}).
The columns of the inverse are calculated by solving the appropriate linear
system of equations by means of a preconditioned conjugate gradient algorithm.
Since the matrix is not hermitian positive definite, there is no guaranteed 
convergence for the conjugate gradient method. In order to make the algorithm 
more robust, especially as the coupling gets strong, we combine the 
conjugate gradient with a continuation scheme: we invert the matrix 
$\rm 1 - \lambda{\tilde M}{\tilde G_0}$ and gradually switch on the 
parameter $\lambda$ from zero to one. The diagonal of the matrix is used as
a preconditioner.

Since upon doing the averages over all of the field configurations one restores
translation invariance and other symmetries, the number of linear systems
of equations that one must solve is much lower than the order of the matrix.

We have checked that the random fields $\rm \vec M$ were sampled properly by 
calculating the two-point correlation function and comparing to the expected
answer Eq.~(\ref{2ptCorr}). A further check was to calculate the second order
term in the expansion Eq.~(\ref{pert}) by Monte Carlo sampling and compare with 
the computation of the appropriate Feynman diagram. The results agreed within
the (small) statistical uncertainty. The conjugate gradient matrix inversion 
was tested on a small system (4 by 4 lattice) at high temperature where it 
could be compared to the result of Gaussian elimination with partial pivoting.

\section{Results}

We shall present results for a doping n=0.8 at a temperature
$\rm T = 200^\circ K$ for a couple of coupling constants $\rm g^2$ and compare 
the results of the computer simulations to lowest order perturbation theory 
(i.e. one spin-fluctuation exchange Feynman graph with bare Green's function) 
and self-consistent one-loop calculations (i.e single spin-fluctuation exchange
Feynman diagram with dressed Green's function). The computations were carried 
out on a $\rm 16\times 16$ lattice with 40 fermion and 21 boson Matsubara 
frequencies. With such parameters, the Green's function, Eq.~(\ref{Green}) 
is obtained by inverting the 20480 by 20480 matrix 
$\rm 1 - {\tilde M}{\tilde G_0}$ for each configuration of the random vector 
fields $\rm \vec M(q)$.

The Eliahsberg renormalization factor $\rm Z(\vec p,i\pi T)$ for a coupling 
constant $\rm g^2 = 0.1\; eV^2$ is shown in fig.(3). The coupling is 
intermediate, since there is a difference between the simple first-order 
perturbation theoretic calculation and the self-consistent one-loop result.
Note that if one tries to calculate the Green's function, Eq.~(\ref{Green})
for each configuration of the random field $\rm \vec M$ by expanding the
denominator in powers of $\rm \vec M$ (using the geometric series), one
finds that the expansion diverges for most field configurations. We have
independently checked that the spectral radius of the matrix 
$\rm {\tilde M}{\tilde G_0}$ is larger than one, but none of its 
eigenvalues are equal to one, so the inverse in Eq.~(\ref{Green}) is
well defined. In this coupling constant regime, the self-consistent 
calculation agrees quite well with the results of the Monte Carlo calculation. 
As expected, it is near the hot spots that the first-order 
perturbation theory and self-consistent calculations deviate from the 
Monte Carlo results. Away from the hot spots, the effective coupling is weak 
(since there is no appreciable difference between the first-order and 
self-consistent calculations) and perturbation theory works satifactorily. 
Near the hot spots, the Monte Carlo simulation yields a renormalization 
factor that is slightly reduced relative to the self-consistent Eliashberg 
result. This must be attributed to third and higher order processes, 
since the lowest order vertex correction enhances the effect of the 
interaction for quasiparticles near the hot spots (Chubukov et. al (1997), 
Monthoux (1997)).

The Eliahsberg renormalization factor $\rm Z(\vec p,i\pi T)$ for a 
coupling constant $\rm g^2 = 0.4\; eV^2$ is shown in fig.(4). While near
cold spots the results of both the first-order perturbation theoretic
and the self-consistent calculations agree with the Monte Carlo
data and each other, one notes that near the hot spots the perturbation
theoretic calculation does a much better job than the Eliashberg 
self-consistent one. The vertex corrections must thus "undress" the
Eliashberg Green's function for this value of the coupling constant. 
Since the results of the simulations near the hot spots yield  a Z some 
20\% larger than the Eliashberg calculation, one can conclude that the 
higher order vertex corrections enhance the spin-fluctuation interaction 
around the antiferromagnetic wavevector when $\rm g^2 = 0.4\; eV^2$. 
We have just begun the exploration of the parameter space and it will 
be interesting to see what happens as all of the parameters (coupling 
constant, antiferromagnetic correlation length and characteristic 
spin-fluctuation frequency) are varied.

\section{Outlook}

We have presented a non-perturbative method for studying the problem 
of quasiparticles coupled to spin-fluctuations and compared our results
to first-order perturbation theory and one-loop self-consistent 
calculations. We have shown some of the limitations of both these 
approaches. We shall study the behaviour of the hot quasiparticles 
as the antiferromagnetic length increases and shall look for the 
pseudo-gap phenomenon seen in ARPES experiments (Loeser et al.,1996),
(Ding et al.,1996). As one approaches the antiferromagnetic transition,
it may not be appropriate to consider only Gaussian terms in the action for
the spin-fluctuations. The method allows the introduction of quartic terms 
in the action of the random magnetic field. The study of the effect of 
such terms on the normal state properties of the system would be of 
great interest. The present formalism can in principle be extended to 
study other Green's functions. We are presently working on such 
extensions.

\section{Acknowlegements}

This work was supported in part by EPSRC. I would like to thank J. Cooper,
J. Loram, G. Lonzarich, C. Nex, D. Pines, D. Scalapino, J. Schmalian for 
helpful discussions on these and related topics.

\begin{figure}
\caption{Spin-fluctuation exchange Feynman graph expansion for the
quasiparticle Green's function (heavy line) obtained by expanding 
Eq.~(\protect\ref{Green}) in powers of the external field $\rm \vec M$.
\label{fig1}}
\end{figure}

\begin{figure}
\caption{Some of the diagrams not included in the expansion. The first 
two RPA-like graphs amount to a renormalization of the spin-fluctuation 
propagator and are already accounted for by our use of the experimental 
$\rm \chi(\vec q,\omega)$. The last two diagrams amount to including 
$\rm (\vec M)^4$ and $\rm (\vec M)^6$ terms in action for the external field
$\rm \vec M$ and are not included in this work.
\label{fig2}}
\end{figure}

\begin{figure}
\caption{Eliashberg renormalization factor $\rm Z(\vec p,i\omega_n) = 1 -
{1\over \omega_n}Im\Sigma(\vec p,i\omega_n)$ at $\rm T = 200^\circ K$
as a function of momentum $\rm \vec p$ for the lowest Matsubara 
frequency $\rm \omega_n = \pi T$. The coupling constant is 
$\rm g^2 = 0.1\; eV^2$.
\label{fig3}}
\end{figure}

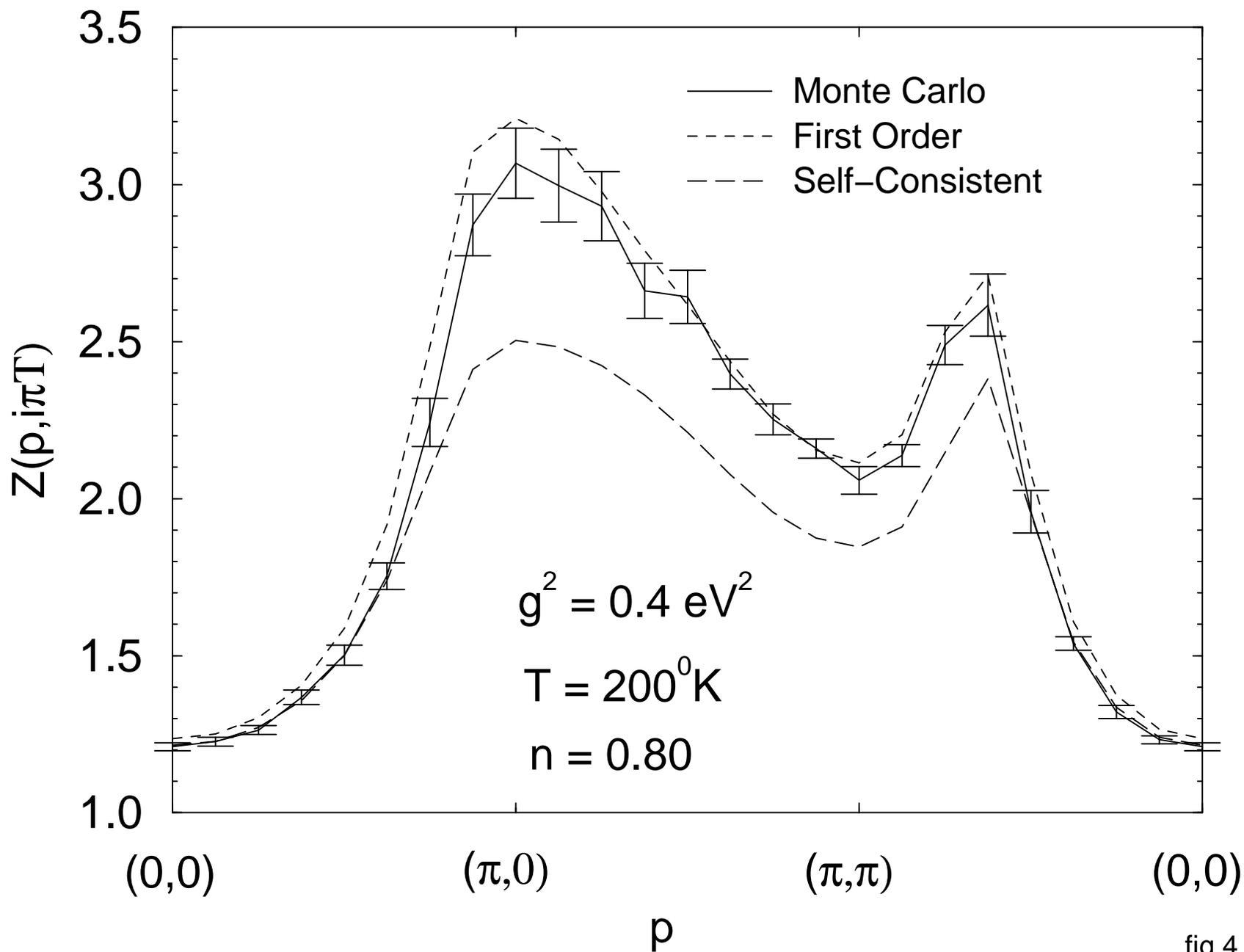
\begin{figure}
\caption{Eliashberg renormalization factor $\rm Z(\vec p,i\omega_n) = 1 -
{1\over \omega_n}Im\Sigma(\vec p,i\omega_n)$ at $\rm T = 200^\circ K$
as a function of momentum $\rm \vec p$ for the lowest Matsubara 
frequency $\rm \omega_n = \pi T$. The coupling constant is 
$\rm g^2 = 0.4\; eV^2$.
\label{fig4}}
\end{figure}


\begin{references}

\noindent A.V. Chubukov, P. Monthoux and D.R. Morr, 
Phys. Rev. B {\bf 56} 7789 (1997).\hfil\break

\noindent H. Ding et al., Nature {\bf 382}, 51 (1996).\hfil\break

\noindent R. Hlubina and T.M. Rice, Phys. Rev. B {\bf 51}, 9253 (1995),
Phys. Rev. B {\bf 52}, 13042 (1995).\hfil\break

\noindent D. Loeser et al., Science {\bf 273}, 325 (1996).\hfil\break

\noindent A.J. Millis, H. Monien and D. Pines, Phys. Rev. B {\bf 42}, 
167 (1990).\hfil\break

\noindent P. Monthoux and D. Pines, Phys. Rev. B {\bf 47}, 6069 (1993).
\hfil\break

\noindent P. Monthoux and D. Pines, Phys. Rev. B {\bf 48}, 4261 (1994).
\hfil\break

\noindent P. Monthoux, Phys. Rev. B {\bf 55}, 15261 (1997).\hfil\break

\noindent J. Schmalian, D. Pines and B. Stojkovic, to appear in Phys. Rev. 
Lett. and unpublished (1998).\hfil\break
 
\noindent J. R. Schrieffer,  J. Low. Temp. Phys. {\bf 99}, 397 (1995).
\hfil\break

\noindent Q. Si, Y. Zha, K. Levin, and J.P. Lu, Phys. Rev. B {\bf 47}, 
9055 (1993).\hfil\break

\end{references}
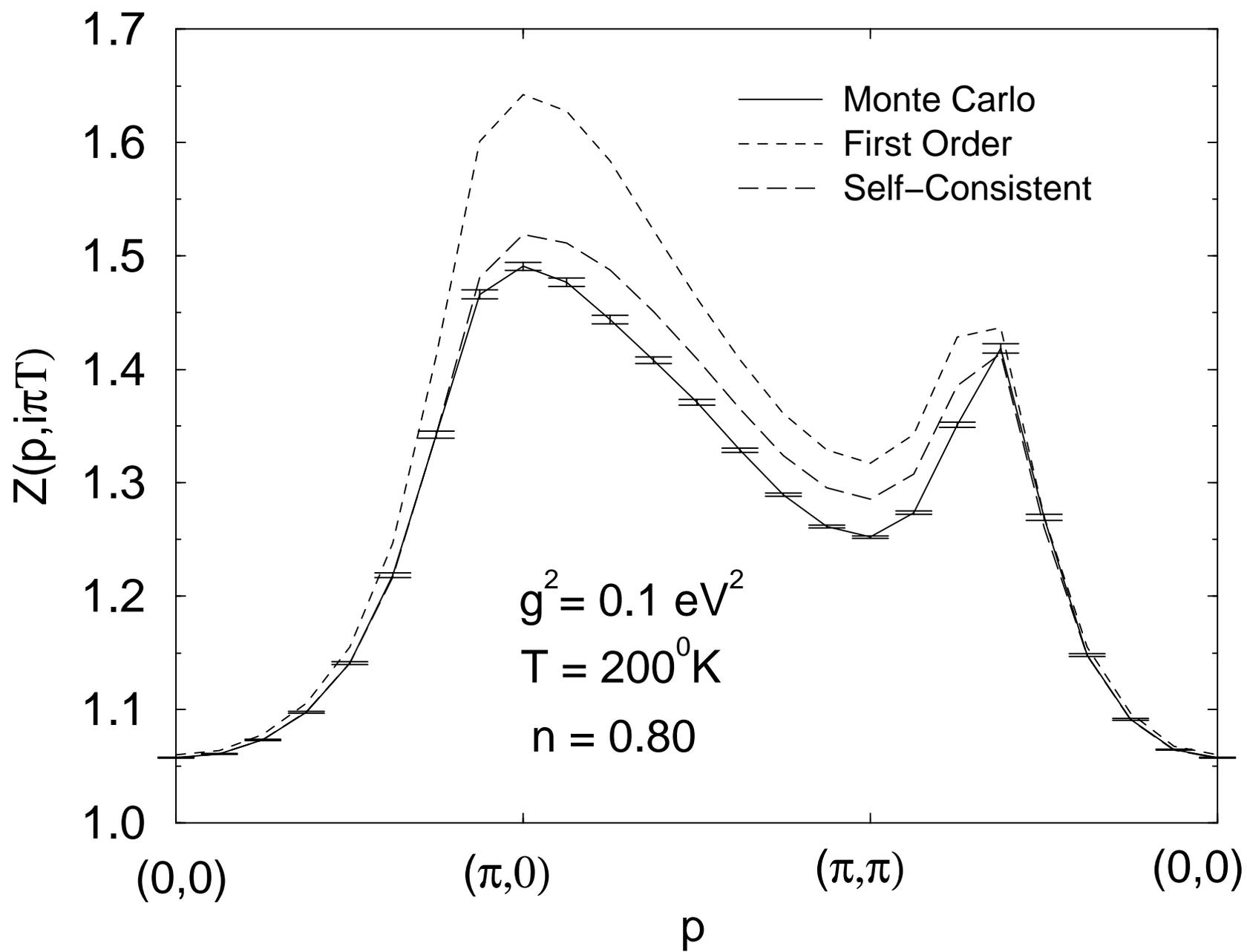
\end{document}